\documentclass[12pt]{iopart}

\usepackage{iopams}

\usepackage{graphicx}
\usepackage{subfigure}%
\usepackage{dcolumn}
\usepackage{bm}
\usepackage{hyperref}
\usepackage{float}
\usepackage{threeparttable}%
\usepackage{amssymb}
\usepackage{amsgen}
\usepackage{amsfonts}
\usepackage{amsbsy}

\hypersetup{
     colorlinks   = true,
     citecolor    = blue
}

\hypersetup{colorlinks=true, citecolor=blue, linkcolor = blue, urlcolor = blue}

\begin{document}

\title[]{The \textit{q}-metric naked singularity: A viable explanation for the nature of the central object in the Milky Way}

\author{F. D. Lora-Clavijo$^{\dag}$, G. D. Prada-Méndez$^{\dag}$, L. M. Becerra$^{\dag}$  and E. A. Becerra-Vergara$^{\dag,*}$}
\address{$^{\dag}$ Grupo de Investigación en Relatividad y Gravitación, Escuela de Física, Universidad Industrial de Santander, A. A. 678, Bucaramanga 680002, Colombia}
\ead{eduar.becerra@correo.uis.edu.co}

\vspace{10pt}
\begin{indented}
\item[]October 2023
\end{indented}

\begin{abstract}
In this work, we investigate whether the compact object at the center of the Milky Way is a naked singularity described by the \textit{q}-metric spacetime.  Our fitting of the astrometric and spectroscopic data for the S2 star implies that similarly to the Schwarzschild black hole, the \textit{q}-metric naked singularity offers a satisfactory fit to the observed measurements.
Additionally, it is shown that the shadow produced by the naked singularity is consistent with the shadow observed by the Event Horizon Telescope collaboration for Sgr-A*. It is worth mentioning that the spatial distribution of the S-stars favors the notion that the compact object at the center of our Galaxy can be described by an almost static spacetime. Based on these findings, the \textit{q}-metric naked singularity turns up as a compelling candidate for further investigation.
\end{abstract}

%
\vspace{2pc}

\vspace{10pc}
\begin{scriptsize}
    $^*$ Author to whom any correspondence should be addressed.
\end{scriptsize}
\maketitle
%
%

\section{\label{sec:Intro} Introduction}

The study of compact objects, dating back to the early theoretical predictions of General Relativity and subsequent astronomical observations of high-energy phenomena, has been central to the development of astrophysics. This pursuit has led to the identification of distinctive signatures in the Galactic Center of the Milky Way associated with a supermassive compact object labeled as Sagittarius-A* (Sgr-A*),  widely believed to be a black hole (BH).

Observations focusing on the gravitational effects that are exerted on light and matter in the vicinity of this compact object have played a crucial role, since their intrinsic nature renders direct measurements unattainable. These observations reveal two different categories that shed light on the properties of Sgr-A*. First, there is the image of the shadow cast by this object. To study this phenomenon, the development of Very-Long-Baseline Interferometry (VLBI)  was necessary,  which allowed the study of the radio spectrum of Sgr-A* on the mm wavelength scale \cite{2008Natur.455...78D}. The Event Horizon Telescope (EHT) collaboration used this technique in an Earth-diameter-scale interferometer to obtain the first image of Sgr-A*'s shadow \cite{2022ApJ...930L..12E}.

Alternatively, the monitoring of stellar orbits in the vicinity of Sgr-A* provides another means to constrain its properties. The UCLA Galactic Center group took the first steps in this direction by characterizing the Keplerian orbit of the S2 star. 
They constrained the mass of Sgr-A* to about $4.1\times 10^6~M_\odot$ by carefully analyzing the astrometric and spectroscopic measurements over a period of 12 years  \cite{2008ApJ...689.1044G}.  Later, by measuring the gravitational redshift \cite{2018A&A...615L..15G} and orbital precession \cite{2020A&A...636L...5G} of the S2 star, the GRAVITY Collaboration further validated and extended our insights into the properties of Sgr-A*.

Drawing from the results of these comprehensive studies, the prevailing theoretical framework for characterizing the nature of the compact object located in the Galactic Center has predominantly centered on that of a Kerr BH.  However, this hypothesis faces challenges from other observations. For instance, the post-pericenter radial velocity of the G2 orbit is slower than the velocity predicted by the geodesic motion in the BH spacetime \cite{2017ApJ...840...50P,2019ApJ...871..126G}. Furthermore, discrepancies arise when comparing the BH spin parameter derived from a recent study by EHT Collaboration \textit{et al.} \cite{2022ApJ...930L..12E} with the constraints established by the S-stars data \cite{2020ApJ...901L..32F,2022ApJ...932L..17F}. 

Therefore, alternative models, commonly referred to as BH mimickers, cannot be easily ruled out \cite{2019ApJ...875L...5E,2022ApJ...930L..17E}. One interesting model is that of naked singularities. These singularities are designated as  `naked' to set them apart from their BH counterparts, which are characterized by the presence of an event horizon. The concept of naked singularity is controversial because it challenges the cosmic censorship hypothesis. Nevertheless, it has been shown that under physically plausible conditions, it is possible to collapse a fluid into a compact object with a curvature singularity, but without an event horizon \cite{1991PhRvL..66..994S}. 

In this work, we focus on the \textit{q}-metric, among the various spacetimes that describe a naked singularity. This choice is based on its status as the simplest static and axisymmetric solution derived from Einstein's equations with a non-vanishing mass quadrupole \cite{2011IJMPD..20.1779Q}. Previous investigations have been conducted to evaluate the model's viability. For example, Arrieta-Villamizar \textit{et al.} \cite{2021CQGra..38a5008A} examined the shadow cast by this particular entity and Prada-Méndez \textit{et al.} \cite{2023CQGra..40s5011P}  explored the emission spectra emanating from geometrically thick accretion disks surrounding it.

The organization of this paper is as follows: in Sec. \ref{qmet} we provide a general description of the \textit{q}-metric, its multipole moments, and the effective relativistic potential associated with this spacetime. Then, in Sec. \ref{datafit} we present the astrometric and spectroscopic measurements fit for the S2 star assuming a naked singularity placed at the Galactic Center while Sec. \ref{shadow} provides a visual representation of the compact object's shadow in the \textit{q}-metric, which has been accurately calculated using the ray-tracing code, \texttt{OSIRIS} \cite {2022EPJC...82..103V, 2023MNRAS.519.3584V}.  Finally, in Sec. \ref{Discussion}, we discuss the results and present the conclusions of this work. Throughout this paper, we use the $(+,-,-,-)$ signature and geometrized units, for which $G = c = 1$.

\section{q-Metric spacetime}\label{qmet}
Weyl's line element \cite{1917AnP...359..117W}  provides the most general way to describe static and axisymmetric spacetimes. The simplest among all possible solutions in the Weyl class is the well-known Schwarzschild metric \cite{1982blho.book...19S}. In terms of its multipole structure, this metric only contains a mass monopole moment. Generalization of this solution can be archived through a Zipoy-Voorhees transformation \cite{1966JMP.....7.1137Z,1970PhRvD...2.2119V}, resulting in a family of solutions with a non-vanishing quadrupole moment. Among these solutions, the so-called \textit{q}-metric stands out for being an exact solution with a handy analytic form and a straightforward multipole physical interpretation \cite{2011IJMPD..20.1779Q}. The line element associated with the \textit{q}-metric  is 

\begin{equation}\label{tensor}\fl
    \eqalign{
    ds^2= & \left(1-\frac{2 m}{r}\right)^{1+q} \mathrm{d} t^2 -\left(1-\frac{2 m}{r}\right)^{-q} \\
    & \Biggl\{ \left(1+\frac{m^{2} \sin ^{2} \theta}{r^{2}-2 m r}\right)^{-q(2+q)}\biggl[\mathrm{d} r^2  \left(1-\frac{2 m}{r}\right)^{-1}  +r^{2} \mathrm{d} \theta^2 \biggr]+r^{2}\sin^{2} \theta \mathrm{d} \varphi^2 \Biggr\},}
\end{equation}


\noindent where $m$ and $q$ are free parameters. It is clear that for $q = 0$, one recovers Schwarzschild's spacetime and for the limiting case $m=0$, Minkowski's spacetime can be recovered by means of a coordinate transformation \cite{2011IJMPD..20.1779Q}.
The physical interpretation of  $m$ and $q$ parameters can be specified by means of the multipole moments, $M_i$: 
%
\begin{equation}
    M_{0}=(1+q) m, \qquad M_{2}=-\frac{m^{3}}{3} q(1+q)(2+q).
    \label{multipole}
\end{equation}
These two terms completely describe the multipole structure of the metric, as higher order moments can be rewritten in terms of them \cite{2011IJMPD..20.1779Q}. The former has been identified as equivalent to the ADM mass \cite{2016PhRvD..93b4024B}, while the latter has been related to the deviation from the spherical symmetry of the source. Thus, this metric describes the outer field of a deformed mass distribution. 
Positive values of $M_{2}$ ($q < 0$) result in a prolate deformation, while negative values of $M_{2}$ ($q > 0$) lead to an oblate deformation.  To obtain positive values of the total mass $M_{0}$, it is generally desirable to have $q>-1$.

An analysis of the Kretschmann scalar, $K = R_{\mu \nu \lambda \tau}R^{\mu \nu \lambda \tau}$, shows that the \textit{q}-metric exhibits curvature singularities at $r = 0$  for all values of $q$, and at the hypersurface $ r = 2m$ for not vanishing value of $q$. Moreover, within the radius of $r = 2m$, two additional singularities appear for values of the parameter $q$ in the range $(-1,-1+\sqrt{3/2})$, excluding zero \cite{2011IJMPD..20.1779Q}.  Luongo and Quevedo \cite{2014PhRvD..90h4032L} have shown that no additional horizon exists outside of the latter hypersurface, thus these singularities correspond to a naked singularity. 

Naked singularities are of great physical interest since they appear to violate the  Penrose cosmic censorship hypothesis\cite{2002GReGr..34.1141P}, which states that every curvature singularity must be surrounded by an event horizon. However,  according to several studies, the outcome of a continuous gravitational collapse can be either a BH or a naked singularity \cite{2007gcss.book.....J}, further supporting the conceivable existence of these compact objects. Moreover, since the existence of a shadow does not imply the presence of a BH,  it is important to remark that the naked singularity of the q-metric cast a shadow, although not all naked singularities do \cite{2019MNRAS.482...52S}. A rather puzzling finding concerning this metric,  reported in the literature, is the identification of regions near the source characterized by apparent repulsive gravitational effects \cite{2021CQGra..38a5008A}.

In particular, for the study of the orbital motion of S2, it is important to perform an analysis of time-like geodesics in the vicinity of the compact object described by the \textit{q}-metric. This approach is justified by the fact that the mass of S2 \cite{2017ApJ...847..120H} is $10^5$ times smaller than the one of Sgr $A^{*}$ \cite{2019A&A...625L..10G}.  The orbital motion of a test particle in the vicinity of this metric is characterized by two constants of motion: the specific angular momentum, $l$, and the specific energy, $E$. These constants are linked to the Killing vector fields   $\xi_{t} = \partial_{t}$ and $\xi_{\phi} = \partial_{\phi}$, respectively.  Thus, from the analysis of geodesic motion in this spacetime \cite{2016PhRvD..93b4024B} we find the following expression 

\begin{equation}
    \left(1+\frac{m^2 \sin ^2 \theta}{r^2-2 m r}\right)^{-q(2+q)} \dot{r}^2={E}^2-\Phi^2,
\end{equation}

\noindent where $\Phi^2$  is the effective relativistic potential given by:

\begin{equation}\fl
    \eqalign{
\Phi^2= & \left(1-\frac{2 m}{r}\right)^{1+q} \\
& \left[r^2\left(1-\frac{2 m}{r}\right)^{-q}\left(1+\frac{m^2 \sin ^2 \theta}{r^2-2 m r}\right)^{-q(2+q)} \dot{\theta}^2\right. \left.+\frac{{l}^2}{r^2 \sin ^2 \theta}\left(1-\frac{2 m}{r}\right)^q+\epsilon\right],}
\end{equation}

\noindent where $\epsilon = 1$ corresponds to time-like particles and $\epsilon = 0$ to null-like particles.

Other studies have evaluated the impact of the deformation parameter $q$ in different phenomena such as geodesic motion and the formation 
of thick and thin accretion disks around the naked singularity described by this metric \cite{2016PhRvD..93b4024B, 2012PhRvD..85j4031C,2003CQGra..20.5121K,2005GReGr..37.1371H,1981PhRvD..24..320P,2021A&A...654A.100F}.

\section{The S2 astrometric data fit}\label{datafit}

The motion of S-stars around Sgr--A* has been the most effective method to explore the nature of the supermassive compact object at the Galactic Center. The most important S-star, S2,  describes a nearly elliptical orbit with an orbital period of around 16 years and a pericentre of approximately 1500 Schwarzschild radius \cite{2003ApJ...586L.127G, 2017ApJ...837...30G, 2018A&A...615L..15G}. The monitoring over the last 20 years of the S2 orbit around the Galactic Center provides the most accurate constraints on the gravitational potential of Sgr--A* to date \cite{2009ApJ...707L.114G,2017ApJ...837...30G,2008ApJ...689.1044G}.

 We solve the full general relativistic motion equations of a test particle in the spacetime geometry generated by the \textit{q}-metric (Eq.~\ref{tensor}) to obtain its real orbit ($r$ vs $\varphi$) and line-of-sight radial velocity (redshift function $z$)
 . The observed orbit ($X_{obs}$ vs $Y_{obs}$) and observed redshift ($z_{obs}$) are associated with the projection of the real orbit onto the sky plane ($X$ vs $Y$) as:
%
\begin{equation}
\eqalign{
    & X_{obs}(t) = X[r(t),\varphi(t);\omega,i,\Omega]+X_0, \\
    & Y_{obs}(t) = Y[r(t),\varphi(t);\omega,i,\Omega]+Y_0, \\
    & z_{obs}(t)= z[r(t),\varphi(t),\dot{r}(t),\dot{\varphi}(t);\omega,i],
}
\end{equation}
where the orbital elements $\omega$, $i$, and $\Omega$ are the argument of pericenter, the inclination between the real and observed orbit, and the ascending node angle, respectively (see Appendix C in \cite{2020A&A...641A..34B} for a graph description of $\omega$, $i$, $\Omega$). The observed position of the star's apparent orbit is defined by the observed angular positions, declination $\delta$ and right ascension $\alpha$,  as
\begin{equation}
\eqalign{
     X_{obs}= & R(\alpha-\alpha_{SgrA^*}), \\
     Y_{obs}= & R(\delta-\delta_{SgrA^*}),
}
\end{equation}
%
with $R=8.3$~kpc the distance from the Earth to the Galactic Center. Finally, the theoretical apparent orbit and radial velocity can be obtained from the real orbit positions through 
\begin{equation}
\eqalign{
     & X= xB+yG, \\
     & Y= xA+yF, \\
     & Z= xC+yH, \\
     & z= \gamma(1+V_z)-1,
}
\end{equation}
%
with $x=r\cos\varphi$ and $y=r\sin\varphi$  and  $B,G,A,F,C$ and $H$,  the known Thiele-Innes constants \cite{2019Sci...365..664D,2020A&A...641A..34B} while $V_z=dZ/dt$ and $\gamma$ is the Lorentz factor. The Constant position offsets $X_0$ and $Y_0$ are introduced to account for the relative position of the gravitational center of mass to the reference frame (for further detail about the real and apparent orbit see \cite{2020A&A...641A..34B}, and references therein).

\begin{figure}[h]
	\centering%
	\includegraphics[width=1\hsize,clip]{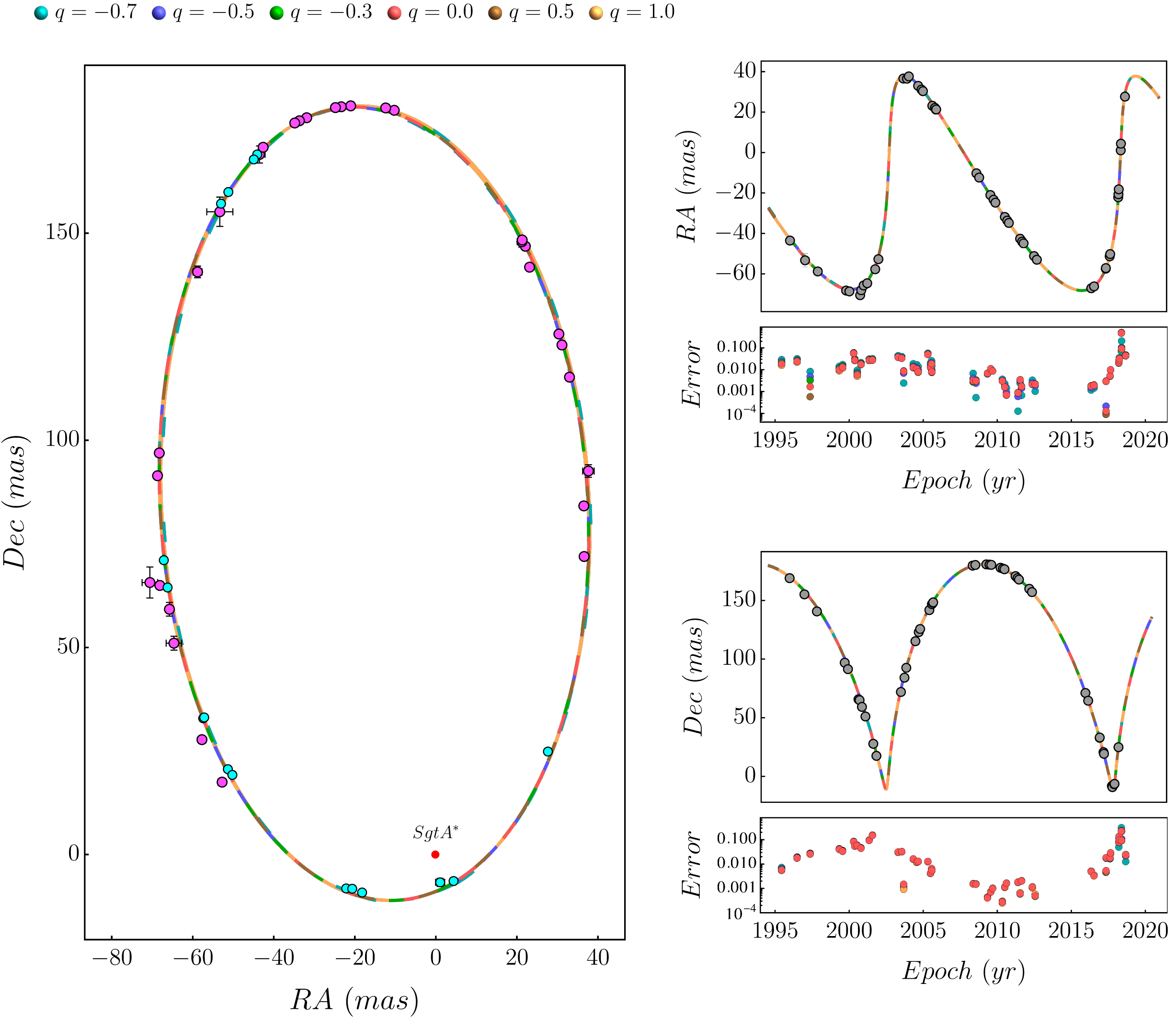}
	\caption{Best-Fitting of the astrometric observations of S2 in orbit around Sgr-A* for different $q$ values (colored lines). \textit{Left panel}: Theoretical and observed position of apparent orbit ($RA$ vs $Dec$). The theoretical star trajectory is shown from 1994.3 to 2026.41 covering two consecutive orbits. Magenta and Cyan observational data correspond to the first and second orbits respectively. The center of mass placed at the origin of coordinates (0,0) is marked by the red point. \textit{Right panel}: $RA$ (top) and $Dec$ (bottom) position as a function of time with their respective errors. We used astrometric measurements from 1995–2018 taken from Do \textit{et al.} \cite{2019Sci...365..664D}.}
	\label{fig:Orbit}%
\end{figure}

\begin{figure}
  \centering
  \includegraphics[width=0.7\textwidth]{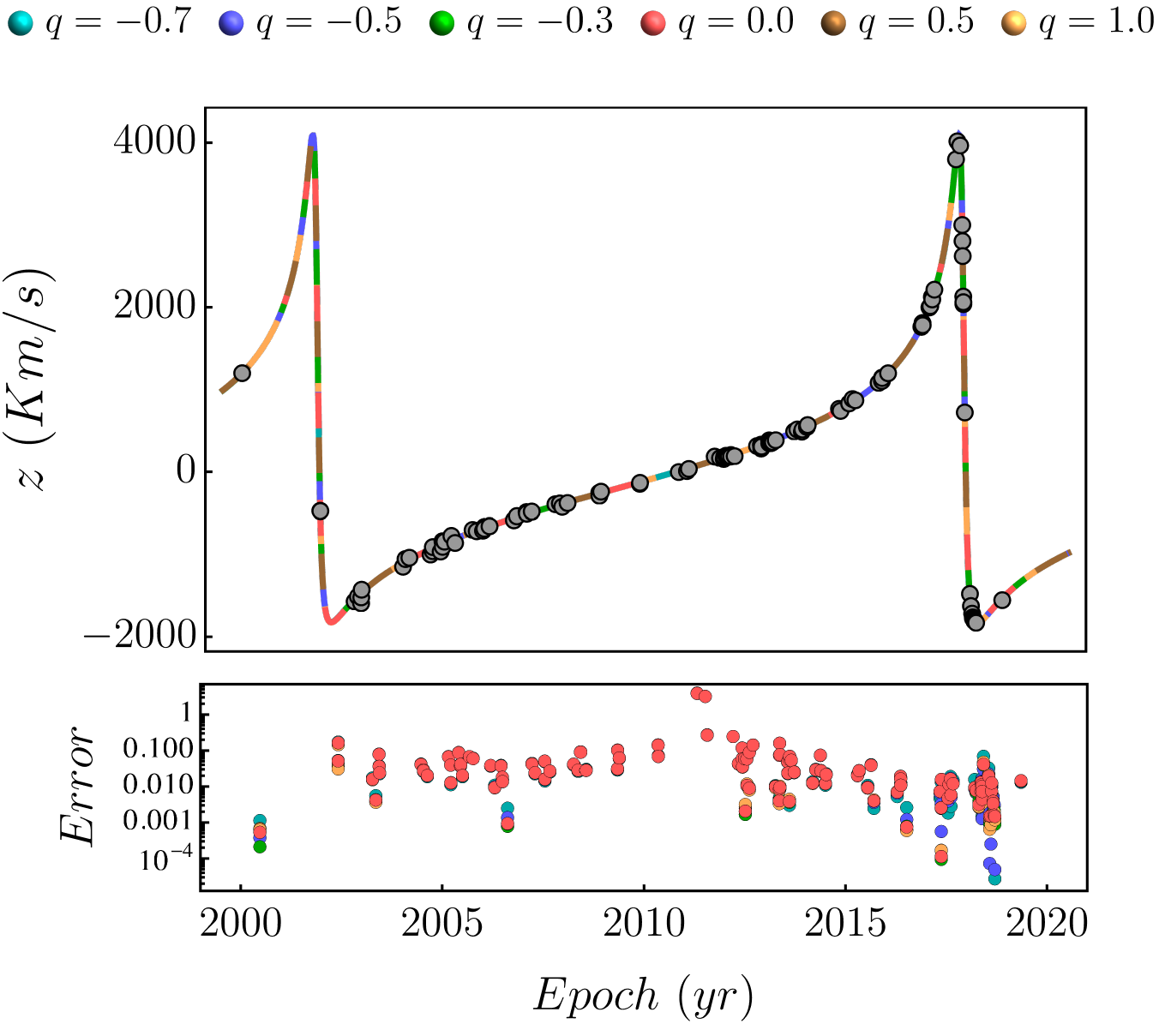}
  \caption{Observed and theoretical radial velocity of S2 for different $q$ values (colored lines). The top panel shows the spectroscopic measurements from 2000–2018, overlaid with our best-fitting velocity on the plane of the sky while the bottom panel shows the error between the observations and predicted velocity. The astrometric observations are taken from Do \textit{et al.} \cite{2019Sci...365..664D}. }
  \label{fig:Velocity}
\end{figure}

We follow the procedure described in \cite{2020A&A...641A..34B}  to obtain the best-fitting parameters of the S2 star orbit around the naked singularity described by the q-metric.
The resulting orbital dynamics 
 are shown in Figures ~\ref{fig:Orbit} and ~\ref{fig:Velocity}. In Figure~\ref{fig:Orbit}, we present the best-fit trajectory of S2's orbit under the gravitational influence of a naked singularity, overlaid with the astrometric measurements. Figure~\ref{fig:Velocity} displays the line-of-sight velocity of the star, calculated in full general relativity for a naked singularity.

The estimated values for the best-fitting parameters and their corresponding reduced-chi-square ($\bar{\chi}^2$) are in Table \ref{tab:parameters}. The theoretical model exhibits a remarkably accurate fit with respect to the observational data for all values of the $q$ parameter used here. This accuracy becomes evident through the error calculated since is obtained that the average deviation hovers around 2\%, 3\%, and 5\%  for the orbital position and radial velocity, respectively (see Figs.~\ref{fig:Orbit} and~\ref{fig:Velocity}). Also, it has been observed that the $\bar{\chi}^2$ displayed by the naked singularity deviates by approximately 10\% in the case of the least optimal fit ($q=-0.7$ and $\bar{\chi}^2\approx 3.81$). Nevertheless, a 1.5\% enhancement is achieved in comparison with the  Schwarzschild BH model ($q=0.0$ and $\bar{\chi}^2\approx 3.45$) for the fitting with $q=1.0$ ($\bar{\chi}^2\approx 3.40$). The above error reported is calculated as $|\xi_{\rm obs}-\xi_{\rm fit}|/{\xi_{\rm obs}}$, where $\xi$ can represent radial velocity as well as the coordinate $X$ or $Y$ of the orbit projected on to the sky plane which we denote as $RA$ and $Dec$, respectively.

\begin{table}[h]
\caption{Summary of the best-fit orbital parameters using the q-metric spacetime for all available astrometric values of S2. For details on the definition of the orbital parameters and on the fitting procedure see \cite{2020A&A...641A..34B}. Note that we did not consider values of $q$ in the range $(-1,-0.7)$ as they exhibited poor data fit.}\label{tab:parameters}
\centering
\resizebox{\textwidth}{!}{%
\begin{tabular}{cccccccccccccccccccccc}
 &  &  &&&  &&&  &&&  &&&  &&&  &&& &  \\
\hline
 & \multicolumn{2}{c}{\textbf{Parameters}} &  &  & $\bm{q = -0.7}$ &  &  & $\bm{q = -0.5}$ &  &  & $\bm{q = -0.3}$ &  &  & $\bm{q = 0.0}$ &  &  & $\bm{q = 0.5}$ &  &  & $\bm{q = 1.0}$ &\\
 \cline{1-3} \cline{1-3}\cline{6-6}\cline{6-6} \cline{9-9}\cline{9-9} \cline{12-12}\cline{12-12} \cline{15-15}\cline{15-15} \cline{18-18}\cline{18-18}\cline{21-22}\cline{21-22}
 &  &  &&&  &&&  &&&  &&&  &&&  &&&  &  \\
 & $e$ &  &&& $0.88576$ &&& $0.88618$ &&& $0.88627$ &&& $0.88628$ &&& $0.88634$ &&& $0.88640$ &  \\
 & $a$ & (as) &&& $0.12522$ &&& $0.12525$ &&& $0.12525$ &&& $0.12525$ &&& $0.12525$ &&& $0.12527$ &  \\
 & $r_p$ & (as) &&& $0.01431$ &&& $0.01426$ &&& $0.01424$ &&& $0.01424$ &&& $0.01424$ &&& $0.01423$ &  \\
 & $r_a$ & (as) &&& $0.23614$ &&& $0.23624$ &&& $0.23626$ &&& $0.23626$ &&& $0.23626$ &&& $0.23631$ &  \\
 & $\omega$ & $(^{\circ})$ &&& $66.4080$ &&& $66.4394$ &&& $66.4299$ &&& $66.4680$ &&& $66.5059$ &&& $66.5285$ &  \\
 & $i$ & $(^{\circ})$ &&& $134.449$ &&& $134.385$ &&& $134.345$ &&& $134.351$ &&& $134.351$ &&& $134.363$ &  \\
 & $\Omega$ & $(^{\circ})$ &&& $228.055$ &&& $228.010$ &&& $227.950$ &&& $227.965$ &&& $227.983$ &&& $228.002$ &  \\
 & $P$ & (yr) &&& $16.0548$ &&& $16.0516$ &&& $16.0513$ &&& $16.0514$ &&& $16.0511$ &&& $16.0505$ &  \\
 & $t_p$ & (yr) &&& $2018.38$ &&& $2018.38$ &&& $2018.38$ &&& $2018.38$ &&& $2018.38$ &&& $2018.38$ &  \\
 & $m$ & $10^6 M_\odot$ &&& $13.5501$ &&& $8.14050$ &&& $5.81526$ &&& $4.07087$ &&& $2.71410$ &&& $2.03683$ &  \\
 & $M_0$ & $10^6 M_\odot$ &&& $4.06502$ &&& $4.07025$ &&& $4.07068$ &&& $4.07087$ &&& $4.07115$ &&& $4.07365$ &  \\
 & $M_2$ & $10^6 M_\odot$ &&& $226.394$ &&& $67.4316$ &&& $23.4020$ &&& $0.00000$ &&& $-12.4956$ &&& $-16.9002$ &  \\
 & $X_0$ & (as) &&& $0.11932$ &&& $-0.02563$ &&& $-0.07655$ &&& $-0.07308$ &&& $-0.07265$ &&& $-0.06113$ &  \\
 & $Y_0$ & (as) &&& $2.47170$ &&& $2.45148$ &&& $2.47572$ &&& $2.47154$ &&& $2.48005$ &&& $2.43476$ &  \\
 & $\bar{\chi}^2$ &  &&& $3.81433$ &&& $3.54945$ &&& $3.48449$ &&& $3.45116$ &&& $3.41916$ &&& $3.40329$ &  \\ 
 &  &  &&&  &&&  &&&  &&&  &&&  &&& &  \\
 \hline
\end{tabular}
}
\end{table}
 %

\section{Shadow in the q metric spacetime}\label{shadow}

We carried out numerical simulations using \texttt{OSIRIS} ({\bf O}rbits and {\bf S}hadows {\bf I}n {\bf R}elativ{\bf I}stic {\bf S}pace-times) \cite{2022EPJC...82..103V, 2023MNRAS.519.3584V}, a code of our authorship based on the backward ray-tracing algorithm for stationary and axially-symmetric space-times. \texttt{OSIRIS} evolves null geodesics “backward in time” by solving the equations of motion in the Hamiltonian formalism. Our code works based on the image-plane model, an assumption where photons come from the observation screen toward the BH. 
In order to appreciate the impact of the deformation on null geodesics close to the naked singularity, we set the numerical integration domain  
 to $x,y \in [-7M_0, 7M_0]$ for $-0.7\le q \le 1.0$. Moreover, the Minkowskian observer is located at  $60^\circ$ from the equatorial plane and a distance of $r = 150M_0$.

\begin{figure*}
\centering
\includegraphics[width=1\textwidth]{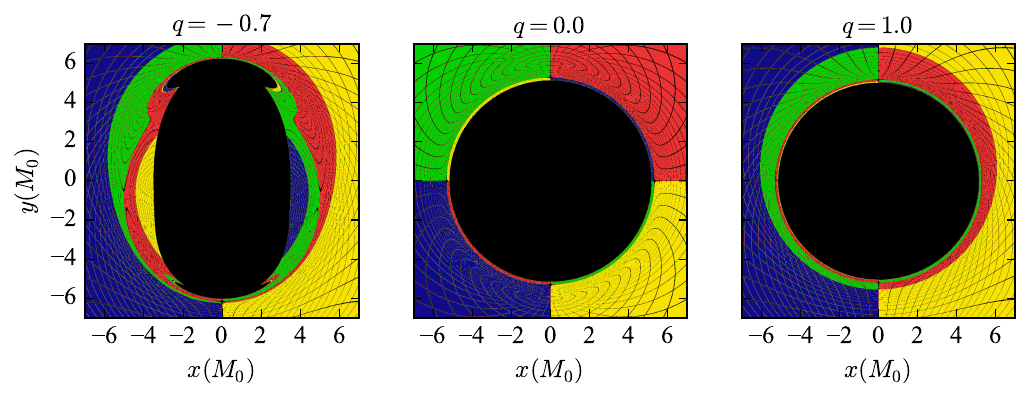}
\caption{Shadow produced by the \textit{q}-metric naked singularity, viewed by a Minkowskian observer located at $60^\circ$ from the equatorial plane, at a distance of $r = 150 M_0$. To classify the orbits, and to elucidate the effect of gravity on null geodesics, we define a celestial sphere as a bright source from which light rays will be emitted. This sphere is concentric with the naked singularity, surrounds the observer, and is divided into four colors: blue, green, yellow, and red. Furthermore, the notion of curvature is provided by a black mesh with constant latitude and longitude lines, separated by $6^\circ$ (see \cite{2022EPJC...82..103V} for more details of the orbits classification).}
\label{fig:lensing}
\end{figure*}

Figure \ref{fig:lensing}  shows the shadow and the gravitational lensing in the vicinity of a naked singularity described by the \textit{q}-metric spacetime for three representative values of the $q$ parameter. As can be observed, the shadow is distorted for $q$ values other than zero. Notably, when it takes negative values, the deformation is prolate, while for positive values, the deformation is oblate  (see \cite{2021CQGra..38a5008A} for a complete analysis of the shadow, Einstein ring, and the gravitational lensing produced by a naked singularity).

In \cite{2022ApJ...930L..12E} was found that only two models described by the Kerr metric satisfy all but the variability constraints imposed by the shadow measurements. Both models have a prograde rotation with dimensionless spines $a=0.5$ and $a=0.94$ and both viewed at an angle of $60^\circ$ with respect to the equatorial plane. 
However, in Fig.~\ref{fig:lensing} we show how shadow values similar to those observed in Sgr-A* can be found with a naked singularity described by the spacetime given by the \textit{q}-metric. 

On the other hand, in \cite{2020ApJ...901L..32F,2022ApJ...932L..17F}, the authors constrain the spin of Sgr A* to $a \le 0.1$, based on the spatial distribution of the S-stars. This result stands in contrast to the findings of \cite{2022ApJ...930L..12E} and favors the notion that the compact object at the Galactic Center is a slowly rotating entity, possibly described by a static spacetime.

Based on all this, the naked singularity described by the \textit{q}-metric not only fits the observed data for the S2 star, but is also consistent with the shadow observed in Sgr-A* and with the notion that its spin is almost negligible.

\section{Discussion and conclusions}\label{Discussion}


In this paper, we examine whether the object at the Milky Way's center can be a singularity without an event horizon. We introduce the concept of a naked singularity using the \textit{q}-metric (see Eq.~\ref{tensor}) and compare its predictions with observations around Sgr-A*, particularly focusing on the behavior of the S2 star and the shadow cast by the compact object in our Galactic Center. 

We utilized available observational data from S2 to narrow down the possible values of the $q$ parameter. Our findings indicate that, within the range of $-0.7 < q < 1.0$, a naked singularity can accurately predict the orbital dynamics of the S2 star, showing a level of precision similar to that offered by a Schwarzschild BH (see Figs.~\ref{fig:Orbit} and \ref{fig:Velocity}). Furthermore, by fitting the astrometric data, we estimated a mass for the central object at approximately $4.07\times10^6 M_{\odot}$, a result that falls within the error window reported by the UCLA team \cite{2019Sci...365..664D} and GRAVITY Collaboration \cite{2022A&A...657L..12G}.

On the other hand, we calculate the shadow produced by the naked singularity for the different values of the parameter $q$. Particularly, we find that the values for the shadow are in agreement with those reported by the EHT collaboration for the compact object in Sgr-A* \cite{2022ApJ...930L..12E}. Something important to highlight here is that the naked singularity described by the $q$-metric corresponds to a static spacetime, which also favors the idea that the central object in our galaxy corresponds to a slowly rotating object \cite{2020ApJ...901L..32F,2022ApJ...932L..17F}. 

Our results not only indicate that a naked singularity cannot be ruled out, but also reinforce the idea that it could be one of the best possible BH mimickers, as suggested by the EHT collaboration \cite{2022ApJ...930L..17E}. This work provides robust observational support to the singularity naked hypothesis as an alternative to the massive BH at Sgr-A*.


\section*{Data availability statements}
The astrometric data used in this work were obtained from Do \textit{et al.}~\cite{2019Sci...365..664D}, and the data here generated are available in Table~\ref{tab:parameters}.
\ack{F.D.L-C was supported by the Vicerrectoría de Investigación y Extensión - Universidad Industrial de Santander, under Grant No. 3703. L.~M.~B is supported by the Vicerrector\'ia de Investigaci\'on y Extensi\'on - Universidad Industrial de Santander Postdoctoral Fellowship Program No. 2023000359. E.~A.~B-V is supported by the Vicerrector\'ia de Investigaci\'on y Extensi\'on - Universidad Industrial de Santander Postdoctoral Fellowship Program No. 2023000354.}



\section*{References}
\providecommand{\newblock}{}

\end{document}